\documentclass[journal]{IEEEtran}

\usepackage{listings}

\usepackage{graphicx}
\usepackage{multirow} 
\usepackage{array}
\usepackage{overpic}

\usepackage{hyperref}
\hypersetup{hypertex=true,
            colorlinks=true,
            linkcolor=blue,
            anchorcolor=blue,
            citecolor=blue}
\usepackage{tcolorbox}

\usepackage{amsmath} 
\usepackage{amssymb}
\usepackage[T1]{fontenc}

\begin{document}

\title{LIVE: LaTex Interactive Visual Editing with NeRF}

\author{Jinwei~Lin,~\IEEEmembership{Member,~IEEE,}
        
\thanks{Jinwei Lin is come from Monash University,} 
\thanks{ORCID of Jinwei Lin:0000 0003 0558 6699,} 
\thanks{Manuscript received May 01, 2023; revised May 29, 2023.}} 

\markboth{Journal or Conference of \LaTeX\, No.~1, May~2023}
{Shell \MakeLowercase{\textit{et al.}}: Simple  Arrow Area Architecture Template}
\maketitle


\begin{abstract}
LaTex coding is one of the main methods of writing an academic paper. When writing a paper, abundant proper visual or graphic components will represent more information volume than the textual data. However, most of the implementation of LaTex graphic items are designed as static items that have some weaknesses in representing more informative figures or tables with an interactive reading experience. To address this problem, we propose LIVE, a novel design methods idea to design interactive LaTex graphic items. To make a lucid representation of the main idea of LIVE, we designed several novels representing implementations that are interactive and enough explanation for the basic level principles. Using LIVE can design more graphic items, which we call the Gitems, and easily and automatically get the relationship of the mutual application of a specific range of papers, which will add more vitality and performance factors into writing of traditional papers especially the review papers. For vividly representing the functions of LIVE, we use the papers from NeRF as the example reference papers. The code of the implementation project is open source.
\end{abstract}


\begin{IEEEkeywords}
Latex, Coding, Graphic, Interactive, Visual, NeRF, Paper
\end{IEEEkeywords}


\section{Introduction}
LaTex, which is usually presented as \LaTeX\, is used as one of the most popular writing and composing methods for completing the document preparation of a paper. LaTex has been popular for more than twenty years \cite{kopka2003guide}. Using LaTex to write a paper is popular and useful without caring about the format and composing overmuch. LaTex and Word are two main formats and methods of writing a paper \cite{matthews2019craft}. But most of the professional and higher level academic journals or conferences prefer using the LaTex as the preferred encoding and writing method. As a popular composing language,  LaTex is powerful in presenting the abundance of paper contents due to its special syntax and coding methods, which leads to the LaTex has been considered as a markup language \cite{kohlhase2004semantic}.

For the normal presentations for the works target or needs, the slides or other presentation documents prefer the type of dynamic presentation. But in the traditional idea, for the articles or papers writing or publication, the dynamic presenting types of documents will be weaker in remaining objectivity and scientificity. The basic reason is that the static content of papers can be printed into the paper and be saved for a long time due to the special features of wooden paper \cite{kupczak2018impact} and the durable ink \cite{ge2009rewritable}. 

For the 2D documents, a good method to preserve them for a long time is using the non-electronic mediums to record the data or information, like paper, metal and stone. In current times, the most commonly used medium is paper, which will usually be presented as a paper book. For now writing academic papers, a good method to preserve them is using the publication and printing of paper books or collections, which can avoid the risk of collapse from electronic storage media like hard disks. Therefore, due to the static feature of 2D paper publications, the main representing style of current paper writing is based on 2D static, which also influenced the main design style of paper writing by using LaTex. On the other hand, due to the publication of the writing academic papers tends to retain the stringency and scientificity of research results, 2D static and non-interactive design or publication style is the main trend. For current design and composing of academic articles, the $word$ and $latex$ are two main editing formats, and the $pdf$ is the most popular format for the final completed results. For the traditional writing of research papers with LaTex tends to use the static components to compose. Due to the main publication style and the representing feature of wooden paper, 2D static representation style will still remain the dominant share for a long time until the basic style of preserving the academic publication is changed. 

Due to the main result of LaTex composing is $pdf$ format files, the whole structure of LaTex paper is mainly focusing on static editing rather than dynamic, which is corresponding to the main preserve document style of 2D static representation documents. There, our research will focus on the interactive feature of $pdf$ format documents. Because the final document type of the research papers will be preserved with wooden papers, if we want to design a new dynamic and interactive method to represent the information of the research paper, the new dynamic component must not influence the preservation of the research paper. The wooden papers can preserve the whole display effect as same as the dynamic interactive paper that doesn’t activate the dynamic interactive effort. Therefore, the dynamic components with a large dynamic or interactive display effect can not be considered, for the transformation or gradual changing of the color and font size. Even every one component of the paper that has special meaning in shape or color can not be changed by dynamic interactive transformations. 

However, there is still a commonly used component, that is the citation command component $cite{ }$ command. Using the command $cite{ }$ command in LaTex can define a citation of a specific reference. This is the only one explicit interactive component in LaTex on PDF files. Using the $cite{ }$ command can define a citation statement and the PDF file will quickly skip to where the cited reference item locates. This is the basic dynamic interactive function that we can research for more complex and meaningful interactions. The function of quick navigating to where the cited reference item is will help the readers easier get the expected information. Combining with the operation $Ctrl + \gets$ will quickly return to the last location, an operations group of quickly going to and coming back is achieved.

In this research, based on the negotiation feature of the citation component of  PDF and LaTex, we will provide several methods to make more colorful and functional implementations of the interactive graphic LaTex components. We will analyze the more efficient method to design an interactive component of LaTex. Based on this idea, several novel methods that can be used for automatically analyzing the citation relationship of multiple papers and improving the efficiency of design of the interactive components mentioned above will be presented in this paper. We will discuss and analyze the better methods to design the LaTex PDF interactive document components. The code of LIVE is open source in GitHub already and will be released after the paper is accepted.

\section{Literature Review}

In this section, we will discuss the research background of the interactive and dynamic research on LaTex and the PDF, analyze why we select the citation-negotiation component of LaTex and PDF as the basic items. The basic research background of the Neural Radiance Fields (NeRF) will be presented.

\subsection{Interactive LaTex}

LaTex, as a popular method to compose the documents, has not too many advantages in interactive implementations. In the early time, researchers studied the possibility of interactive scientific formulae input \cite{crisanti1986easytex}. Some researchers focused on the interactive implementations with other formats documents \cite{dill1991interactive}. Other researchers were interested in designing the visual graphic applications for the interactive edicting and $Tex $ code generation \cite{schrod1994towards}. During that time, the Tex is the basic format for those documents that are the basic format for the LaTex in the later time. As the technology developed, the current LaTex has been developed to a good level. TCurrent interactive style of Latex is focusing on multiple functional rich text and hypertext in the web design \cite{bouvin2019notecards}. Hoever, the existing research that studies on the interaction features of LeTex with PDF is rarely research on the interaction feature of the native component of LaTex. For the academic and scientific composing with LaTex, the research and implementation of designing novel methods to make LaTex PDF more interaction is significant. 

\subsection{Interactive PDF}

For academic research result writing, the most popular format of the generated documents of LaTex is the PDF. The format type of the PDF document is rich text, which means PDF document can contain multiple types of components, for example, text, tables and figures. If the analysis perspective is only set on an electronic document, a PDF document is powerful in interactive editing and displaying. These documents are usually designed for working and teaching \cite{nolan2007dynamic}, some documents have the interaction features in embedding editing \cite{barnes2013embedding}, which are composed as a digital document \cite{masson2020chameleon}. However, if the purpose of using the LaTex to generate a PDF document is composing an academic paper or article, there are more factors that need to be considered. For example, most of the current published academic articles or papers retain a convention that academic publications should be available to be preserved as physical publications, like the paper book or papers collections or journals. Therefore, for the academic publication PDF documents, too dynamic or interactive functions are not suitable. Generally, the interactive or dynamic functions that can influence the physical publication are not suitable. Therefore we select the citation-negotiation component of LaTex with PDF as the basic design item.

\subsection{Neural Radiance Fields}
Neural Radiance Fields (NeRF) is a novel implementation method of using the 2 dimensional (2D) images to generate a three dimensional (3D) sense or model \cite{mildenhall2021nerf}. In the implicit view synthesis research area, NeRF is a hot topic research subarea. There are abundant numbers of research directions and results toward NeRF. The research area of NeRF can be divided into Dynamic NeRF \cite{pumarola2021d} and Static NeRF \cite{mildenhall2021nerf}.  Classifying by the rendering objects, can be divided into NeRF for multiscale representation \cite{barron2021mip}, NeRF for wild environments \cite{martin2021nerf} and NeRF for human \cite{xu2021h} etc. There are enough examples for LIVE to make the discussion and analysis for the novel interactive methods. Therefore we select the NeRF as the main study case to verify LIVE theory and methods.

\section{Methodology and Analysis}

We propose several novel methods to achieve the implementation of interactive LaTex PDF documents. The basic item is the citation-negotiation component of LaTex with PDF. We call the novel items that are designed and developed based on the citation-negotiation component or other visual interactive item of PDF file that are based on this principle as $Gitem$ which represent the meaning of graphic interactive items. In this secession, we will analyze and discuss the design and operating principle of the $Gitem$s.

\subsection{LIVE Architecture}
\begin{figure}[t]
\centering
\includegraphics[width=0.95\columnwidth]{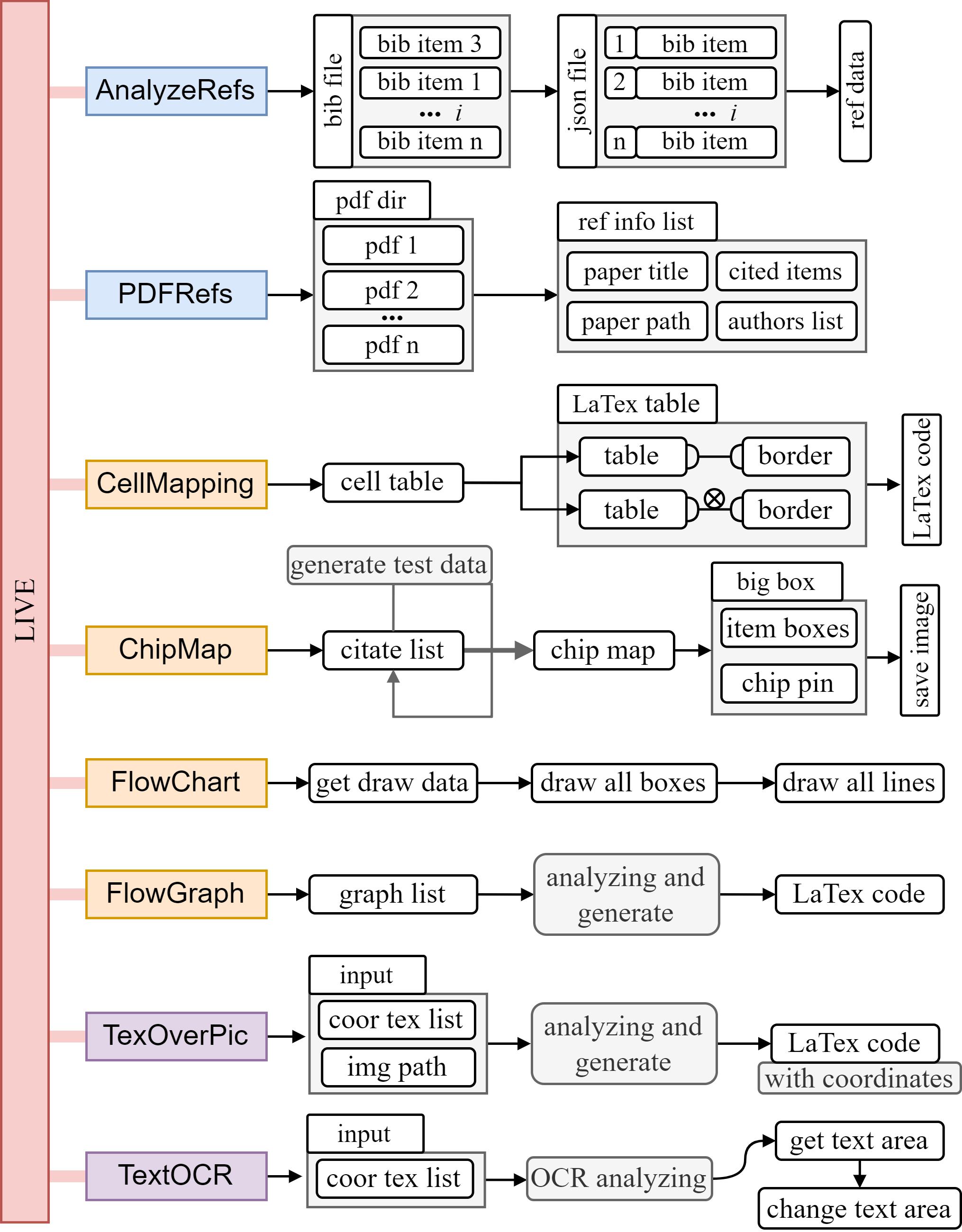}
\caption{Basic Architecture of LIVE.}
\label{fig2}
\end{figure}

As shown in Figure \ref{fig2}, there are 8 main components of LIVE. The $AnalyzeRefs$ is the part that is designed to analyze the $bib$ format file and transform the information of $bib$ file to a json variable, which contains all of the citation information. The $PDFRefs$ part is used to automatically analyze and extract all of the key information of a PDF paper. The analyzed and extracted information items of the paper include: the title of the paper, the path of the paper, the authors list of the paper and all of the cited reference items of the paper. The $AnalyzeRefs$ and $PDFRefs$ are designed for the PDF and reference items analyzing, therefore, this kind of items are classified to Gitems Analyzing Auxiliary (GAA). 

Subsequently, we analyze the Gitems of this version of LIVE. Note that there are four Gitems  will be stated and analyzed in this paper, but this doesn’t mean the maximum of the Gitems is four. Based on the citation-negotiation component principle of LaTex with PDF, more meaningful and multifunctional Gitems can be designed and achieved. Subsequently, the component  $CellMapping$ is a simple Gitem that is used to generate the LaTex table code with a border or not. The component $ChipMap$ is Gitem that can represent the interactive citation relationship between different papers in a specific group. The $generate test data$ is a module of $ChipMap$ that can generate the test data automatically, which is called the $citate list$. The $citate list$ is the input and the output is the $chip map$ which is a citation relationship map graph and is designed based on the principle and shape of chip and pin structure. The generated $chip map$ image will be saved in the preset directory. The basic structure of the $chip map$ is a big box, and many little item boxes are set around the four sides of the big box. Gitem $chip map$ is the most complex Gitem in this version of LIVE. After them, the Gitem $FlowChart$ is a component that can automatically draw a flowchart based on the given flow relationship. Drawing the boxes and lines are two main functions of $FlowChart$. Finally, the Gitem $FlowGraph$ is designed to generate a novel type of interactive graph and automatically get the LaTex code of the $FlowGraph$. More complex design and configuration is also supported. 

The third part of the components is the part called Gitems Processing Auxiliary (GPA), as shown as Figure \ref{fig2}, which includes the $TexOverPic$ and $TextOCR$ two GPAs. The GPA$TexOverPic$ is designed to generate the LaTex code that can write the specific words on the images according to the text coordinates list $coor text list$. The GPA$TextOCR$ is designed to get the coordinates of the texts or words that are in an image, then change the corresponding area with another color, for example the white color.

\subsection{AnalyzeRefs GAA Interface}

\begin{figure}[t]
\centering
\includegraphics[width=0.99\columnwidth]{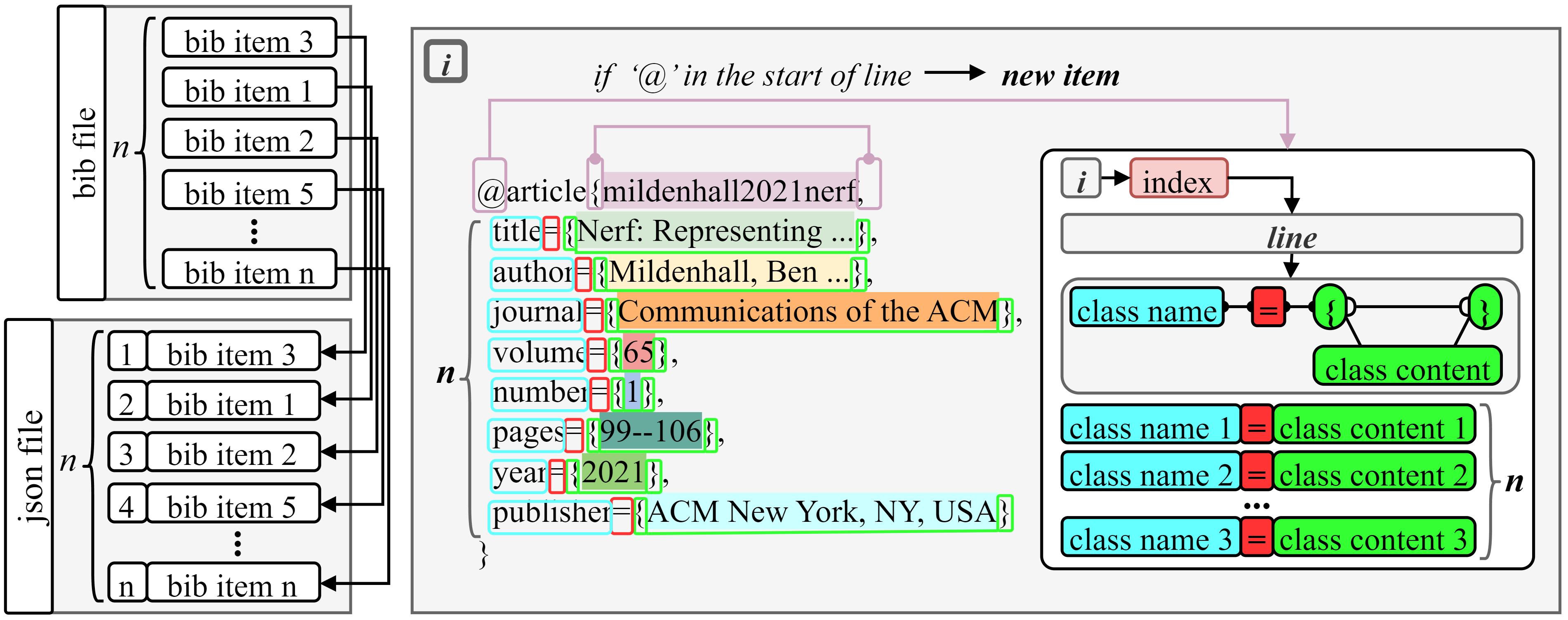}
\caption{AnalyzeRefs Component Design Principle.}
\label{fig3}
\end{figure}

As shown in Figure \ref{fig3}, the main processing of GAA $AnalyzeRefs$ is extracting the specific information from each citation item of the target paper. Here, we selected the citation paper \cite{mildenhall2021nerf} as the study case. Using the GAA $AnalyzeRefs$ can quickly extract the detailed information of each citation item of a specific paper.

As shown in Figure \ref{fig3}, the sorting sequence of the citation items of a paper is not always the same as the sequence of how the citation items appear in the context of the paper, which means the sort of the items in the $bib$ file is often different from the sort of the citation number in the paper. According to the structure feature of a json file, a main key for each item in the json file that includes the extracted data from the $bib$ file should be individual and meaningful. For this issue, one method is to set the title or citation label of each citation item as the main key of the json item. Gaining the title or citation label of a citation item from the $bib$ file can reduce computing procedures when making the searching and analyzing of the target item. However, this method will result in a risk of different json items having the same main key, which is conflit. Meanwhile, in designing the programming, selecting the title or citation label as the main key of each json item will be more unsuitable and difficult than selecting the index $i$ as the main key. 

As shown in Figure \ref{fig3}, each item of the $bib$ file will be extracted and transformed to the corresponding json item in the generated json file. In the detailed processing, reading the $bib$ file by lines one by one. Subsequently, analyzing each individual line text. If line text includes the $@$ character, it means the current is the start line of the lines of an unitary bib item. Using the function of gaining the index of a specific character in a string variable to get the indexes of left brace $\{$ and right brace $\}$, as well as the equal sign $=$, according to the fixed sorting sequence, to get the citation label and the class name and corresponding class content of each json item. 

The output of the processing of GPA $AnalyzeRefs$ Interface is a json file that includes the detailed citation information of a target paper or $bib$ file, which can be used by other components.

To test the performance of GAA $AnalyzeRefs$, we select the part of $bib$ file of this paper as input. As shown in Code \ref{code1}, it is a part of output when using the GAA $AnalyzeRefs$ to analyze the $bib$ file of this paper. The output result includes the $type$, $title$, $author$ and other items that are stated in the $bib$ file for each cited paper. 

\begin{lstlisting}[language={Python}, caption={Code 1}, label={code1}]
bid_D = {1: {'type': 'article', 'title': 'Craft beautiful equations in Word with LaTeX', 'author': 'Matthews, David and others', 'journal': 'Nature', 'volume': '570', 'number': '7760', 'pages': '263--264', 'year': '2019', 'publisher': 'Nature'}, ...}
\end{lstlisting}

\subsection{PDFRefs GAA Interface}


GAA $PDFRefs$ is designed to analyze the information of each citation paper item in a paper. In the code of the $PDFRefs$ module of LIVE, the function $process\_all\_refs$ is the main function of the processing of GAA $PDFRefs$. For better code management, each GAA of LIVE is developed separately in a Python file and implemented by being designed as a class.

\begin{figure}[t]
\centering
\includegraphics[width=0.99\columnwidth]{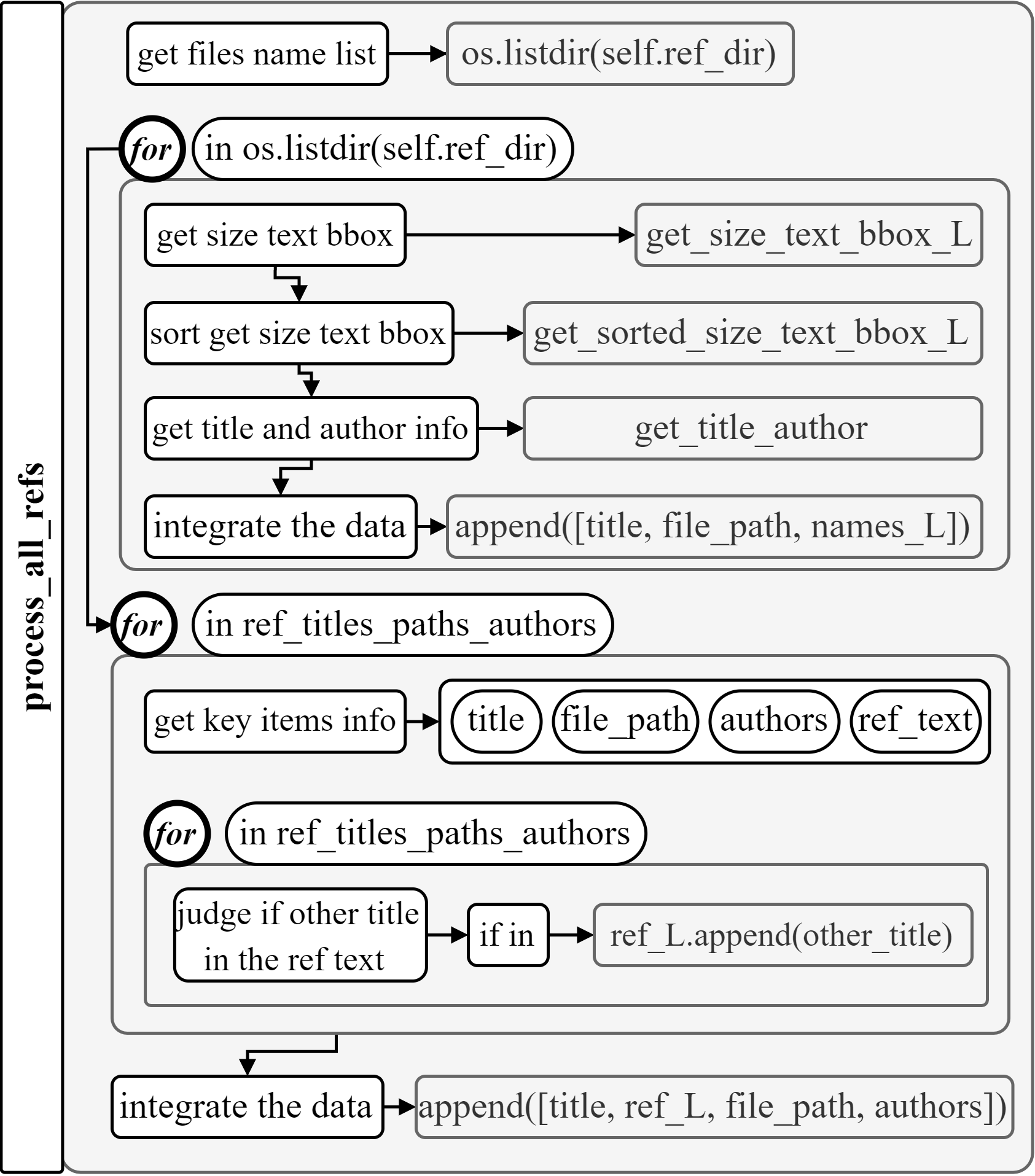}
\caption{PDFRefs Component Design Principle.}
\label{fig4}
\end{figure}

During all programming of the project of LIVE, we use the detailed naming method to name the variables and functions as well as classes, which will include the reading and editing conveniences. To specifically analyze the core of GAA $PDFRefs$, we design the Figure \ref{fig4} that includes the main components. 


As shown in Figure \ref{fig4}, the first step is getting the list variable that includes the names of all files in the targeted directory that preserves all of the pdf files that are planned to be analyzed. Using the $os.listdir$ function to achieve it. Subsequently, enter a for-loop processing for each pdf file in the pdf preserving directory. Here the first sub-process is to get the size, text and $bbox$ information of each page of the analyzed pdf file, with the using of the $fitz$ library that is a Python library to extracting the text or image content of a pdf file following a special format. The second step is to sort the size, text and $bbox$ information list with the function $get\_sorted\_size\_text\_bbox\_L$. Following by using the function $get\_title\_author$ to get the title and the author list of the analyzed paper. When analyzing whether a noun is a person name or not, we design the function $if\_a\_name$ that judges whether a noun is a prison by analyzing the numbers and locations of the lowercase or uppercase. The third step is integrating each group of the items $title$, $file\_path$ and $names\_L$ as a unit item into the record list variable. Subsequently, enter the second stage processing, which is exactly entering the second for-loop process. The traversed object is the $ref\_titles\_paths\_authors$ list from the output of the first for-loop process. In this process, the main operation is to get the key items information of each cited reference paper of the analyzed paper, including the items of the $title$, $file_path$, $authors$ authors list and the $ref$ referenced papers list. The most important is using the second sub-for-loop to judge whether other papers of the global $bib$ file are included in the analyzed paper, and record the referenced papers in variable $ref\_L$. Finally integrating all of the gained information and returning the result.

To test the performance of GAA $PDFRefs$, we selected two papers, \cite{muller2022instant} and named it as $Instant-NGP$ and the paper \cite{mildenhall2021nerf} and named it as $NeRF$, as the input of GAA $PDFRefs$, the output is shown as Code \ref{code3}. As shown in Code \ref{code3}, the GAA $PDFRefs$ can analyze and extract the title, author especially the interactive citation items information of each papers in the target directory. From the output, it can be simple to get the interactive citation relationship of the papers in a target directory. For example, paper \cite{muller2022instant} cited the paper \cite{mildenhall2021nerf}, but paper \cite{mildenhall2021nerf} does not cite paper \cite{muller2022instant}.

\begin{lstlisting}[language={Python}, caption={Code 2}, label={code3}]
ref_titles_refs_paths_authors_L = [['Instant Neural Graphics Primitives with a Multiresolution Hash Encoding', ['NeRF: Representing Scenes as Neural Radiance Fields for View Synthesis'], './references/pdf/Instant-NGP.pdf', ['THOMAS MULLER', 'ALEX EVANS', 'CHRISTOPH SCHIED', 'ALEXANDER KELLER']], ['NeRF: Representing Scenes as Neural Radiance Fields for View Synthesis', [], './references/pdf/NeRF.pdf', ['Ben Mildenhall', 'Pratul P. Srinivasan', 'Matthew Tancik', 'Jonathan T. Barron', 'Ravi Ramamoorthi', 'Ren Ng']]]
\end{lstlisting}

\subsection{CellMapping Gitem Interface}

\begin{figure}[t]
\centering
\includegraphics[width=0.9\columnwidth]{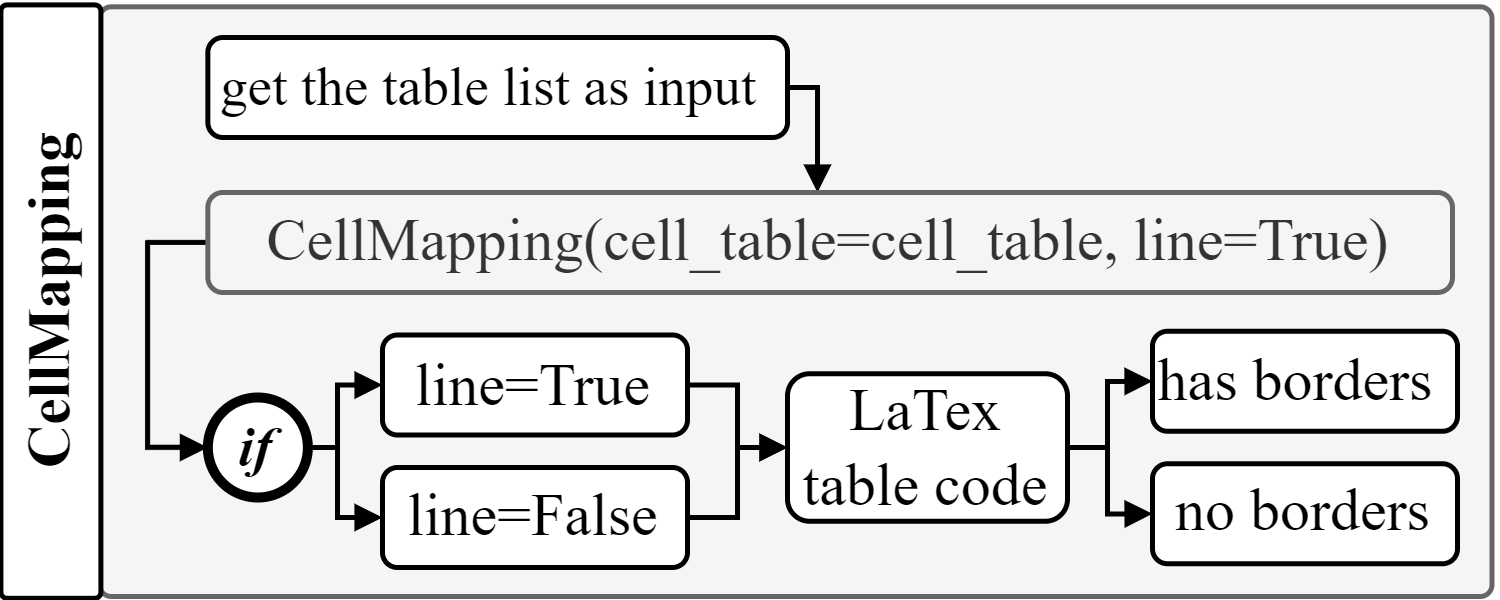}
\caption{CellMapping Gitem Design Principle.}
\label{fig5}
\end{figure}

Gitem $CellMapping$ is designed to generate a LaTex table item code according to the pre-set python list variable. As shown in Figure \ref{fig5}, the input of Gitem $CellMapping$ is the Python list variable that states the structure and content that are corresponding to the  target LaTex table, which will be implemented in class $CellMapping$. There is a customized configuration of setting whether the generated LaTex table has borders or not, which is configuring the bool variable $line$ in the parameters list of class $CellMapping$. 

The Gitem $CellMapping$ is a simple component of LIVE. As shown in Code \ref{code4}, the main processing of Gitem $CellMapping$ is to get the generated LaTex code of the target LaTex table. We use the $T\_$ as the prefix to name the variables to make it more readable and easier to read and develop the Python code. The key code of this part is the double nested for-loop in lines 15 through 26. This process is to generate the matrix items arrangement part of the table LaTex code. After gaining all of the component LaTex code, the next step is to integrate them in one to generate the final output. 

\begin{lstlisting}[language={Python}, caption={Code 3}, label={code4}]
{get self.T_begin,self.T_caption,self.T_label,self.T_renewcommand,self.T_tabcolsep,self.p_centering,begin_tabular,self.p_centering...}
self.T_table_matrix=''
for x in range(self.cell_w_num):
  str_x=hline+''
  for y in range(self.cell_h_num):
    x_y=self.cell_table[x][y]
    if y != self.cell_h_num-1:
      if x_y: str_x+=str(x_y)+' & '
      elif x_y == None: str_x+=' '+' & '
    else:
      if x_y: str_x+=str(x_y)
      elif x_y == None: str_x+=' '
  str_x+='   \\\\'+'\n'
  self.T_table_matrix+=str_x
self.T_table_matrix+=hline+'\n'
\end{lstlisting}

To test the performance of Gitem $CellMapping$, we selected three papers about the NeRF: \cite{mildenhall2021nerf}, \cite{barron2021mip} and \cite{pumarola2021d} as the samples. As shown in Code \ref{code5}, the  relative information of the LaTex table is collected and integrated as the Python 2D list which has the same structure of the target LaTex table. The variable $cell\_table$ will be used as an input of Gitem $CellMapping$ and the output, as shown in Code \ref{code6} is the generated LaTex table code. The user can make some little changes or modifications in that code to make it be more suitable for the target generated object.

\begin{lstlisting}[language={Python}, caption={Code 4}, label={code5}]
cell_table=[
  ['Item','Citation','Year','Movement','Object'],
  ['NeRF',r'\cite{mildenhall2021nerf}','2021','Static','Normal'],
  ['Mip-nerf',r'\cite{barron2021mip}','2021','Static','Multiscale '],
  ['D-nerf',r'\cite{pumarola2021d}','2021','Dynamic','Normal '],
  ['Instant-NGP',r'\cite{muller2022instant}','2022','Static','Normal'],
]
\end{lstlisting}

\begin{lstlisting}[language={TeX}, caption={Code 5}, label={code6}]
...
\begin{tabular}{| p{1.35cm}<{\centering} | p{1.35cm}<{\centering} | p{1.35cm}<{\centering} | p{1.35cm}<{\centering} | p{1.35cm}<{\centering}| }
\hline Item & Citation & Year & Movement & Object   \\
\hline NeRF & \cite{mildenhall2021nerf} & 2021 & Static & Normal   \\
\hline Mip-nerf & \cite{barron2021mip} & 2021 & Static & Multiscale    \\
\hline D-nerf & \cite{pumarola2021d} & 2021 & Dynamic & Normal    \\
...
\end{lstlisting}


After gaining the Code \ref{code6}, just copy and paste the geneated code into the LaTex project, with some additional modifications, then the final result that is shown as Table \ref{table label} is achieved.

\begin{table}\centering
\caption{table title}
\label{table label}
\renewcommand{\arraystretch}{1}
\tabcolsep=0.1cm
\begin{tabular}{| p{1.35cm}<{\centering} | p{1.35cm}<{\centering} | p{1.35cm}<{\centering} | p{1.35cm}<{\centering} | p{1.35cm}<{\centering}| }
\hline Item & Citation & Year & Movement & Object   \\
\hline NeRF & \cite{mildenhall2021nerf} & 2021 & Static & Normal   \\
\hline Mip-nerf & \cite{barron2021mip} & 2021 & Static & Multiscale    \\
\hline D-nerf & \cite{pumarola2021d} & 2021 & Dynamic & Normal    \\
\hline 
\end{tabular}
\end{table}

\subsection{ChipMap Gitem Interface}

Gitem $ChipMap$ is the most complex and powerful Gitem in this version of LIVE. The main function of Gitem $ChipMap$ to generate a chip map that represents the interactive citation relationship between specific or customized different papers.

\begin{figure}[t]
\centering
\includegraphics[width=0.9\columnwidth]{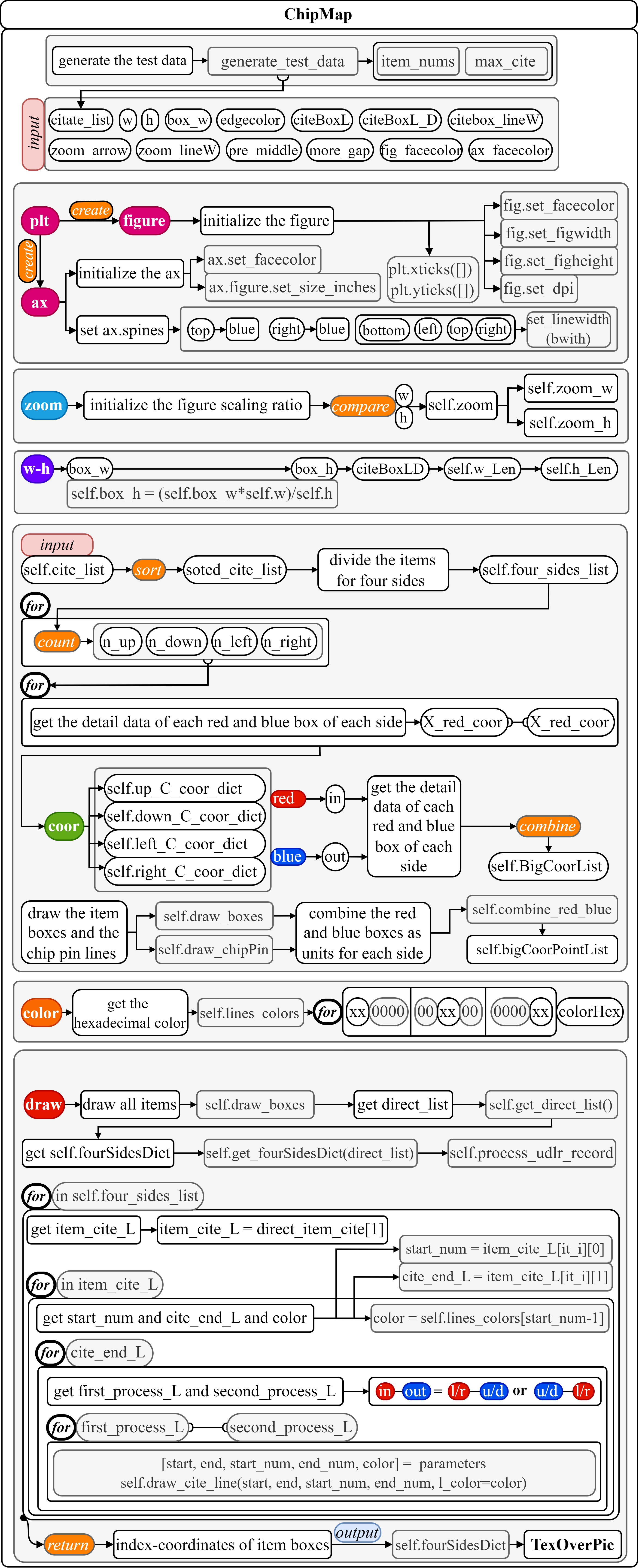}
\caption{CellMapping Gitem Design Principle.}
\label{fig6}
\end{figure}

As shown in Figure \ref{fig6}, the main processing of Gitem $Chipmap$ is implemented in the class $Chipmap$. There are abundant input parameters that make the Gitem $Chipmap$ will more customizable. The most important parameter is $citate\_list$, which describes the interactive citation relationship between the analyzed referenced papers. The  $citate\_list$ follows the data format as $citate\_list=[[ni, [c1, c2, …, cj]]; ni, cj\in \mathbb{N}^+$. Parameter $ni$ represents the index number of the current analyzed paper and the parameter list $ [c1, c2, …, cj]$ represents all of the cited referenced papers of the current analyzed paper $ni$. The variable $citate\_list$ can be gained from the calculation of another module,  or for test, using the function $generate\_test\_data$ to generate, which can be customized by configuring the $item\_nums$ and $max\_cite$.

Subsequently, using the $matplotlib$ library that is a popular and native library of python as the basis of drawing. As shown in Figure \ref{fig6}, creating two main components $figure$ and the $ax$ and making the corresponding configuration for drawing. To guarantee quality of the generated chipmap, making a good modification of the size and width-height aspect ratio of the generated image is important. As shown in Figure \ref{fig6}, for zooming, firstly initializing the scaling ratio of the image with comparing the width and height of the image. Then calculating the relative parameters about the width-height aspect ratio, which are prepared for the final image generating.

Following, enter the processing of the calculation of the coordinates of the boxes of the $ChipMap$. As shown in Figure \ref{fig6}, the $self.cite\_list$ will be sorted In order of the number of citations from largest to smallest as $sorted\_cite\_list$. All of the items in $sorted\_cite\_list$ will be divided into four groups following the directions of: up, down, left and right. Then using a for-loop to calculate the number of the items on each side, followed by getting the detailed data of the red boxes and blue boxes of each side. Using the $X\_red\_coor$ and $X\_blue\_coor$ to represent each red and blue box respectively. Here, the red box means the in and citing direction of the citation flow, while the blue box means the out and being-cited direction of the citation flow. Next, using the $self.BigCoorList$ to combine all of the box's coordinates lists for each side. Eventually, calling the function $self.draw\_boxes$ and $draw\_chipPin$ to draw the citation line of each citation relationship. Next, using the $self.BigCoorPointList$ to rearrange the red and blue boxes coordinates lists for each side for further processing.

Subsequently, handling the color configuration for each citation paper. If all of the citation papers are using the same color to represent the citation relationship, the citation pin lines of the $ChipMap$ must be indistinguishable and confused if the number of citation pin lines is too large. Therefore we use the hexadecimal color value arrangement method to handle this issue. As shown in Figure \ref{fig6}, according to the number of citation items, taking the hexadecimal color values averagely across the hexadecimal color space in $\#xx0000$, $\#00xx00$ and $\#0000xx$, which means averagely allot the colors of citation items into a special $RGB$ units color space. 

Next processing is the drawing. This is also the most important of the Gitem $ChipMap$ is implemented in the function $draw\_al\l_citations$. Firstly gaining the $direct\_list$ that contains the sides information of all of the box items. Subsequently, using this data to get the $self.fourSidesDict$ that contains the side box key information of each side, followed by using the $self.process\_udlr\_record$ to set the initialization of boxes number record for each side. Then using a for-loop for the list item in $self.four\_sides\_list$ which will get the $item\_cite\_L$ first, which contains the detailed citation information of each side, followed by entering the second for-loop in $item\_cite\_L$, which will get the output item index as $start\_num$ and the list that states all of the cited items index of the current item as a input list as $cite\_end\_L$, as well as the color data. This also means that for one analyzed paper item, using a list to contain all of the cited paper items of this analyzed paper item. The number of the cited paper items can be none or more. Following, entering the third for-loop in $cite\_end\_L$ that states the process of gaining the $first\_process\_L$ and  the $second\_process\_L$ that are two main lists which contain two different kinds of chip citation lines which are presented by the citation lines inside the $ChipMap$. If the start point of chip citation lines are left or right and the end point are up or down, or vice versa, then this kind of citation lines should belong to the $second\_process\_L$, and other citation lines will belong to the $first\_process\_L$. The reason why we divided the citation lines into two main kinds and used two methods to handle them sequentially is that, for the citation lines that belong to $first\_process\_L$ are rectangular polygon lines, which are easier to be drawn, and for the citation lines that belong to $second\_process\_L$ are polylines, which need to consider more about the overlap and conflict situations. We design more processes methods for $second\_process\_L$ to address this problem, of which the main key is to set the transfer points of polylines and the offset when overlapping conflicts occur. Finally in the for-loop in $first\_process\_L$ and  $second\_process\_L$ respectively, calling the function $self.draw\_cite\_line$ to make the drawing.

Finally, the final output result is the $self.fourSidesDict$ that contains the index coordinates of item boxes that are distributed on four sides of the big square of Gitem $ChipMap$. Combining with the GPA module $TexOverPic$, then writing the citation text on the corresponding location to generate a LaTex code. Using LaTex code with the $ChipMap$ image can build an interactive Gitem $ChipMap$.

To test the draw performance of the Gitem $ChipMap$, we use the automatically generated data from function $generate\_test\_data$ to test. As shown in the Figure \ref{fig1}, the parameter $item\_nums$ represents the number of the citation papers that are analyzed. The parameter $item\_nums$ can be odd or even number. The parameter $max\_cite$ represents the number of the papers that will cite the current paper, which means there will be 1 to $max\_cite$ papers will cite the current paper. Configuring the $zoom\_arrow$ can set the shape size of the arrow of the citation lines. Configuring the $zoom\_lineW$ can set line width of the citation lines. The boxes set around the four sides of Gitem $ChipMap$ are the $citebox$. Each $citebox$ represents one paper that is analyzed and is corresponding to one little red box and one blue box. The width of the $citebox$ can be customized. As shown in Figure\ref{fig6}, as soon as the size and the resolution of the generated image is enough, the Gitem $Chip$ can support hundreds of analyzed paper items. The whole process will be fast.


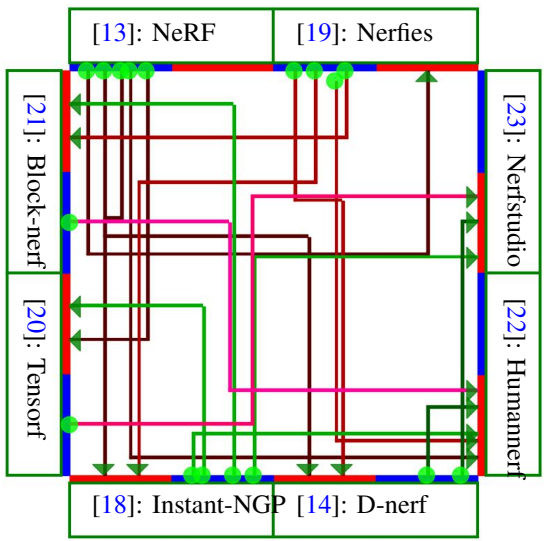
\begin{figure}[htbp]
\begin{overpic}[width=0.4\textwidth]{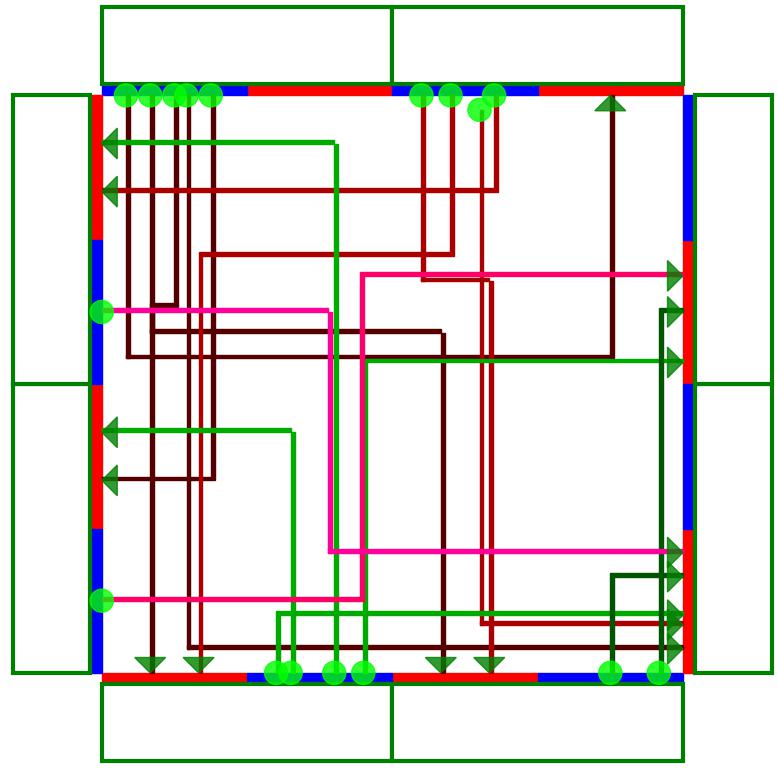}
\put(0.06799999999999999\textwidth, 0.37000000000000005\textwidth){\cite{mildenhall2021nerf}: NeRF}
\put(0.21800000000000003\textwidth, 0.37000000000000005\textwidth){\cite{park2021nerfies}: Nerfies}
\put(0.06799999999999999\textwidth, 0.024\textwidth){\cite{muller2022instant}: Instant-NGP}
\put(0.21800000000000003\textwidth, 0.024\textwidth){\cite{pumarola2021d}: D-nerf}
\put(0.018\textwidth, 0.18200000000000002\textwidth){\rotatebox{-90}{\cite{chen2022tensorf}: Tensorf}}
\put(0.018\textwidth, 0.332\textwidth){\rotatebox{-90}{\cite{tancik2022block}: Block-nerf}}
\put(0.37000000000000005\textwidth, 0.18200000000000002\textwidth){\rotatebox{-90}{\cite{weng2022humannerf}: Humannerf}}
\put(0.37000000000000005\textwidth, 0.332\textwidth){\rotatebox{-90}{\cite{tancik2023nerfstudio}: Nerfstudio}}
\end{overpic}
\caption{Test Gitem ChipMap with NeRF.}
\label{fig7}
\end{figure}

To test the performance of the Gitem $ChipMap$ in practical application, we selected some of the analyzed papers that are relative to NeRF to get the map of their interactive caution relationship. Using the descriptive interactive data from GAA $PDFRefs$ get the interactive table and make some transformations. Subsequently, using the Gitem $ChipMap$ to generate the ChipMap as shown in Figure \ref{fig7}, followed by using the GPA $TexOverPic$ to generate the LaTex code with an important component $overpic$. Finally copying the ChipMap image and the LaTex code into the LaTex project then can get the display of Gitem $ChipMap$.

\subsection{FlowChart Gitem Interface}

\begin{figure}[t]
\centering
\includegraphics[width=0.95\columnwidth]{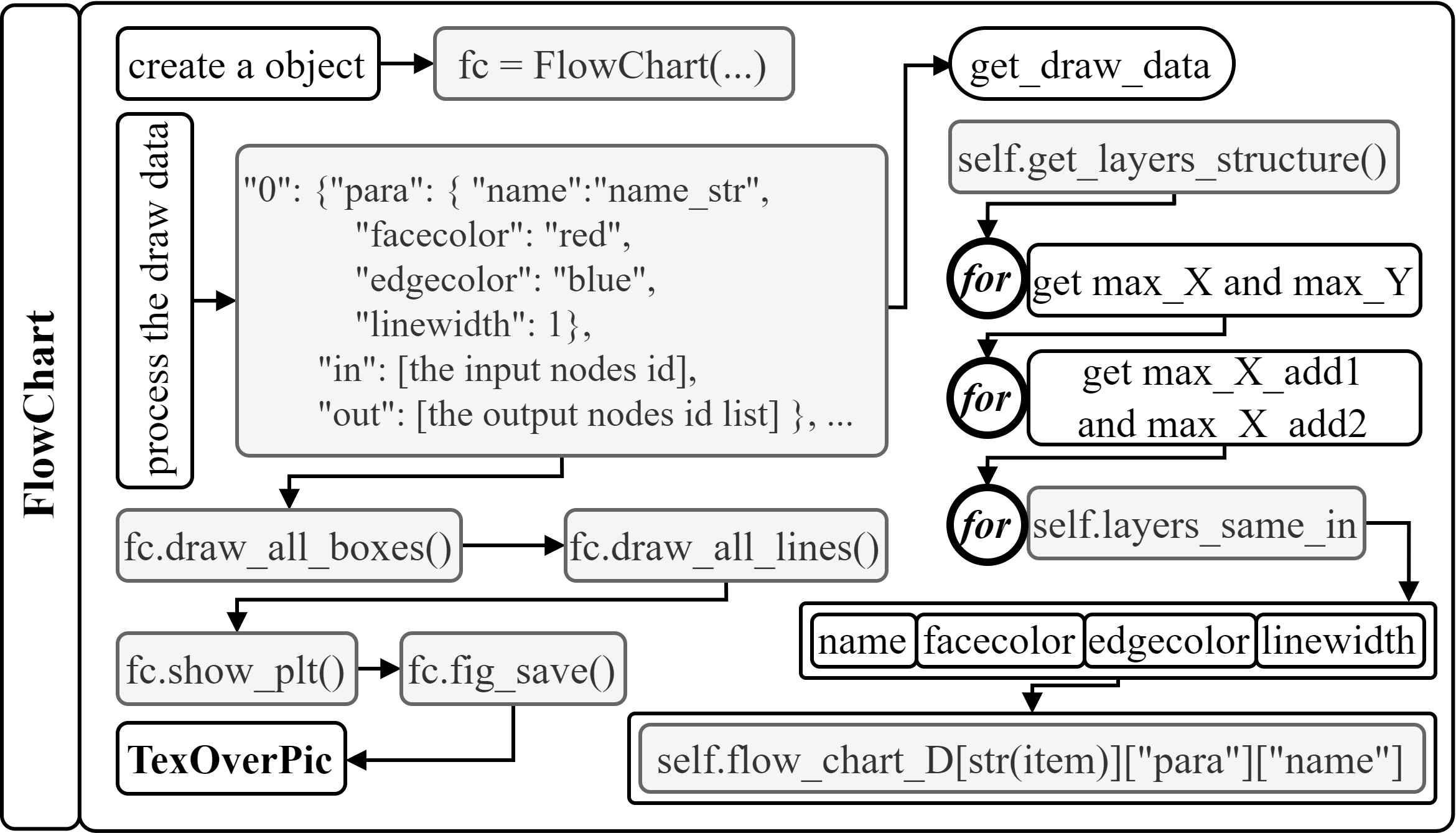}
\caption{FlowGraph Gitem Design Principle.}
\label{fig8}
\end{figure}

\begin{figure}[htbp]
\begin{overpic}[width=0.4\textwidth]{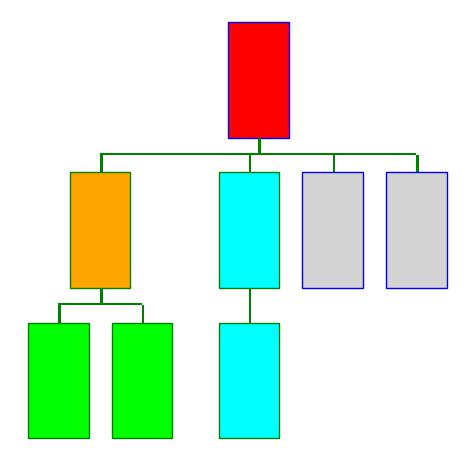}
\put(0.046\textwidth, 0.11000000000000001\textwidth){\small \rotatebox{-90}{\cite{wang2023f2} F2-NeRF}}
\put(0.118\textwidth, 0.11000000000000001\textwidth){\rotatebox{-90}{\cite{li2022nerfacc} Nerfacc}}
\put(0.20999999999999996\textwidth, 0.11000000000000001\textwidth){\rotatebox{-90}{\cite{chan2022efficient} Efficient}}
\put(0.20999999999999996\textwidth, 0.24\textwidth){\rotatebox{-90}{\cite{pumarola2021d} D-nerf}}
\put(0.082\textwidth, 0.24\textwidth){\rotatebox{-90}{\cite{muller2022instant} NGP}}
\put(0.28200000000000003\textwidth, 0.24\textwidth){\rotatebox{-90}{\cite{weng2022humannerf} Human}}
\put(0.35400000000000004\textwidth, 0.24\textwidth){\small \rotatebox{-90}{\cite{garbin2021fastnerf} Fastnerf}}
\put(0.21800000000000003\textwidth, 0.37\textwidth){\rotatebox{-90}{\cite{mildenhall2021nerf} NeRF}}
\end{overpic}
\caption{Test Gitem FlowChart with NeRF.}
\label{fig9}
\end{figure}

The main function of Gitem $FlowChart$ is automatically generating a flow chart with interactive citation items that are located in the corresponding item boxes. As shown in Figure \ref{fig8}, the core of principle of Gitem $FlowChart$ is the function $get\_draw\_data$ that is designed to get the correct format data for the boxes and lines drawing. As shown in Figure \ref{fig9}, to test this Gitem, we select some of NeRF related papers to get the $FlowChart$ result.

\subsection{Horizontal-vertical Analysis Method}
To discuss and analyze the content of this review better, in this subsection, we propose the analyzing method the Horizontal-vertical Analysis Method (HVAM). Detailed, the HVAM is an analyzing method that focuses on two dimensions to make the analysis, the horizontal dimension and vertical dimension. Usually, the horizontal dimension is represented by time, and the vertical dimension is represented by the items in the same range. The items in the same range usually means the items that belong to the same category.

\begin{figure}[t]
\centering
\includegraphics[width=0.95\columnwidth]{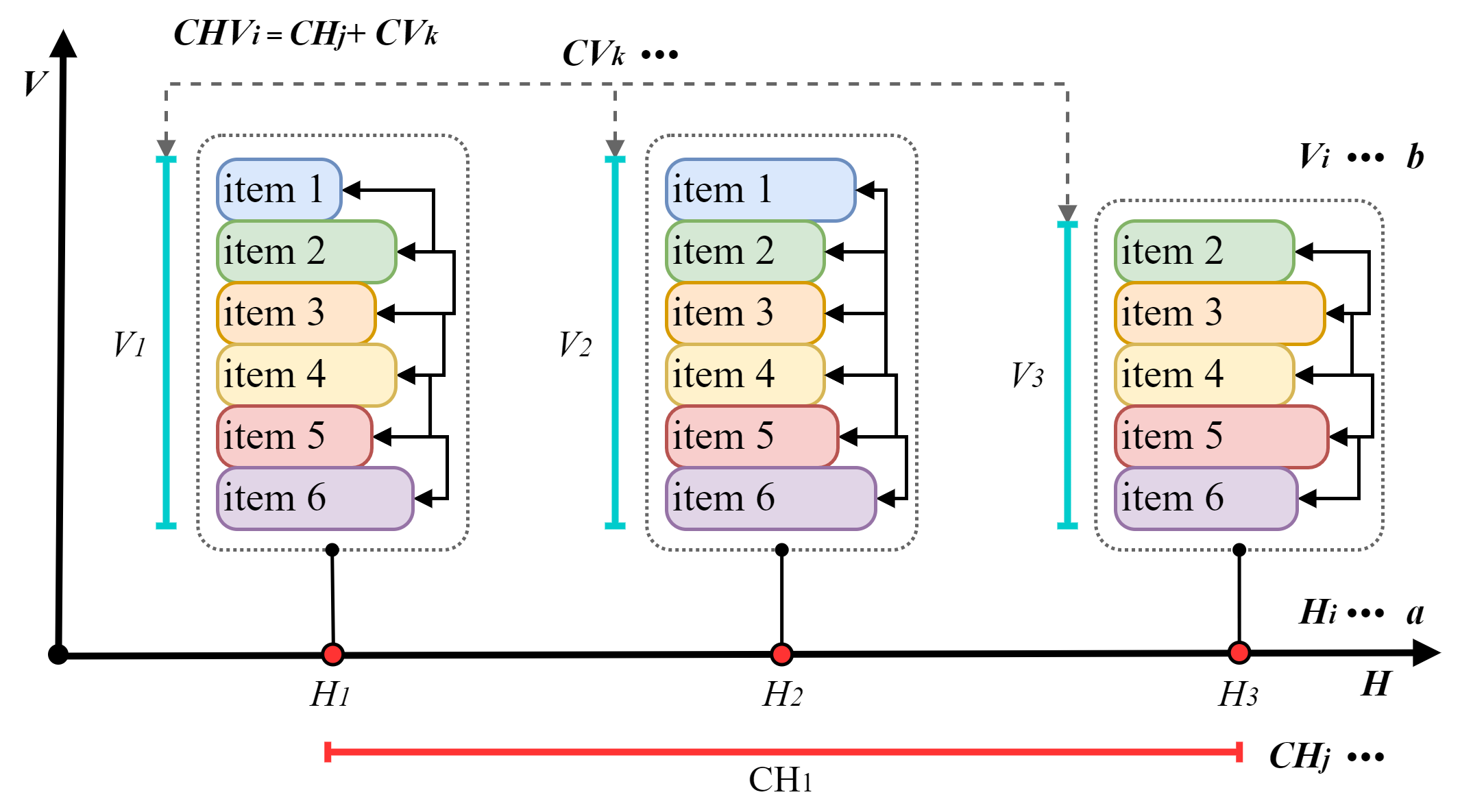}
\caption{Horizontal-vertical Analysis Method Principle.}
\label{fig10}
\end{figure}

As shown in Figure \ref{fig1}, the HVM method can be considered as a two-dimensional (2D) comparison method, which includes the horizontal dimension and vertical dimension. We use parameter $C$ to represent the whole comparison of using the HVAM, which follows the equation $C =  {\textstyle \sum_{i}^{s} CHV_i}; i,s \in \mathbb{N}+$. As shown in Equation \ref{eq1}. The parameter $CHV_i$ represents the $i\text{-}h$ Horizontal-vertical comparison that is combined with the corresponding $j\text{-}h$ horizontal comparison $CH_j$ and the corresponding $k\text{-}h$ vertical comparison $CV_k$. $CH_j$ and $CV_k$ satisfy  $CH_j =  {\textstyle \sum_{i}^{a} H_i}; i,a \in \mathbb{N}+, i \le a$. $CV_k =  {\textstyle \sum_{i}^{b} H_i}; i,b \in \mathbb{N}+, i \le b$. The sum of the horizontal comparisons is $n$ and the sum of the vertical comparisons $m$. The $j\text{-}h$ horizontal comparison $CH_j$ consists of $H_i$ whose sum is $a$, and the $k\text{-}h$ horizontal comparison $CV_k$ consists of $V_i$ whose sum is $b$. Note, the parameter $j$ is not necessarily equal to parameter $k$, and the parameter  $a$ is not necessarily equal to parameter  $b$. This means the number of the horizontal comparison needs not to be equal with the number of the vertical comparison. Usually, for each horizontal comparison $CH_j$, there are more than one vertical comparison $CV_k$.

\begin{equation}
\label{eq1}
CHV_i =  {\textstyle \sum_{j}^{n} CH_j} + {\textstyle \sum_{k}^{m} CV_k}
\end{equation}

The final result of the HVAM is the comprehensive comparison result of $C$, one important characteristic of HVAM is the individual comparison results of $CHV_i$, which can represent the comparison result for each individual stage.

\begin{table}[htbp]\scriptsize
\begin{center}
\caption{Test Gitem FlowGraph with NeRF.}
\label{tb2}
\tabcolsep=0cm
\renewcommand\arraystretch{1.5}
\begin{tabular}{c}

$
\begin{matrix}
& \underbrace{\boxed{2021}}  & \underbrace{\boxed{2022}}  & \underbrace{\boxed{2023}}   &  \\[0pt]
 & \boxed{\star \rightarrow} & \boxed{\star\star \rightarrow} & \boxed{\star\star\star \rightarrow} \\[0pt]
& \uparrow & \uparrow & \uparrow  \\[0pt]

& \boxed{
\begin{matrix}
\begin{tcolorbox}[colback=green, width=2cm, left=0mm, right=0mm, top=0mm, bottom=0mm, boxsep=0.5mm, arc=0mm, boxrule=0pt, bottomrule=0pt, toprule=0pt, title={}]
$
\begin{matrix}
\cite{mildenhall2021nerf}NeRF  \\
\cite{yu2021pixelnerf}pixelnerf  \\
\cite{wang2021nerf}NeRF\text{-}\text{-}  \\
\cite{barron2021mip}Mip\text{-}nerf  \\
\cite{pumarola2021d}D\text{-}nerf  \\
\end{matrix}
$
\end{tcolorbox}\end{matrix}
}

& \boxed{
\begin{matrix}
\\
\\
\begin{tcolorbox}[colback=yellow, width=2cm, left=0mm, right=0mm, top=0mm, bottom=0mm, boxsep=0.5mm, arc=0mm, boxrule=0pt, bottomrule=0pt, toprule=0pt, title={}]
$
\begin{matrix}
\cite{hong2022headnerf}Headnerf  \\
\cite{tancik2022block}Block\text{-}nerf  \\
\cite{barron2022mip}Mip\text{-}nerf 360  \\
\end{matrix}
$
\end{tcolorbox}\end{matrix}
}

& \boxed{
\begin{matrix}
\\
\\
\\
\begin{tcolorbox}[colback=pink, width=2cm, left=0mm, right=0mm, top=0mm, bottom=0mm, boxsep=0.5mm, arc=0mm, boxrule=0pt, bottomrule=0pt, toprule=0pt, title={}]
$
\begin{matrix}
\cite{bian2023nope}Nope\text{-}nerf  \\
\cite{bao2023sine}Sine  \\
\end{matrix}
$
\end{tcolorbox}\end{matrix}
}

\end{matrix}
$
\end{tabular}   
\end{center}   
\end{table}


The main function of Gitem $FlowGraph$ is automatically generating a time sequence or other developing sequence flow interactive graph to describe the development of one issue. As shown in Table \ref{tb2}, we used some of the NeRF paper items to test the effect of Gitem $FlowGraph$. The number of one row can be customized. The year item can be changed to other text and the layout is automatically calculated and generated. The layout supports the word wrap following the number of set items of a line.

\subsection{TexOverPic and TextOCR}

\begin{figure}[t]
\centering
\includegraphics[width=0.95\columnwidth]{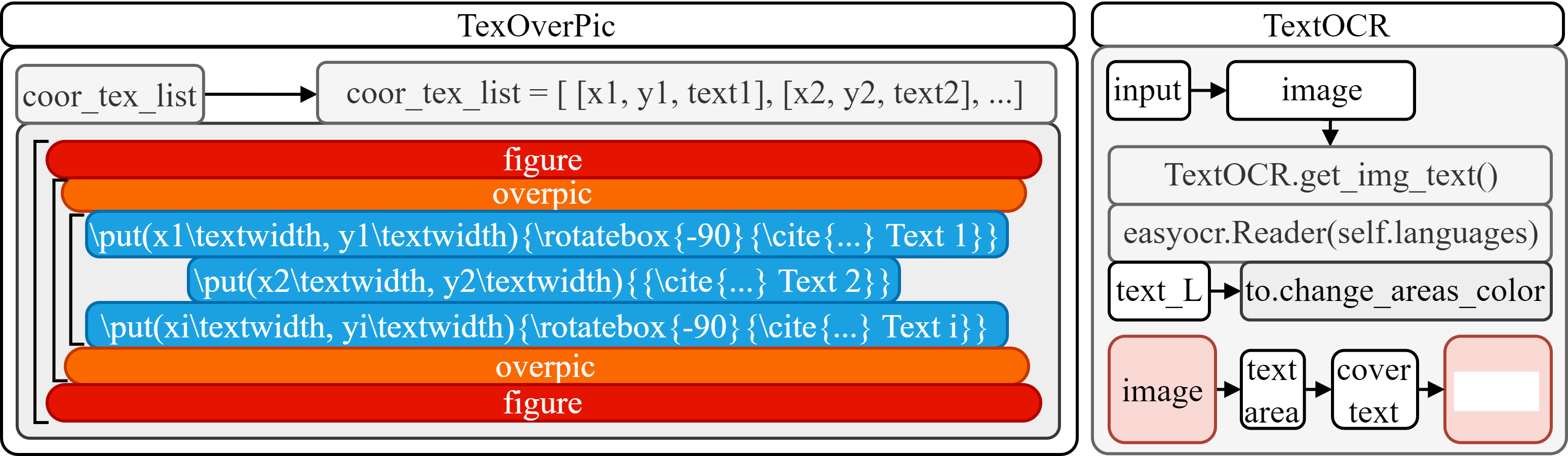}
\caption{TexOverPic and TextOCR Design Principle.}
\label{fig11}
\end{figure}

\begin{figure}[t]
\centering
\includegraphics[width=0.95\columnwidth]{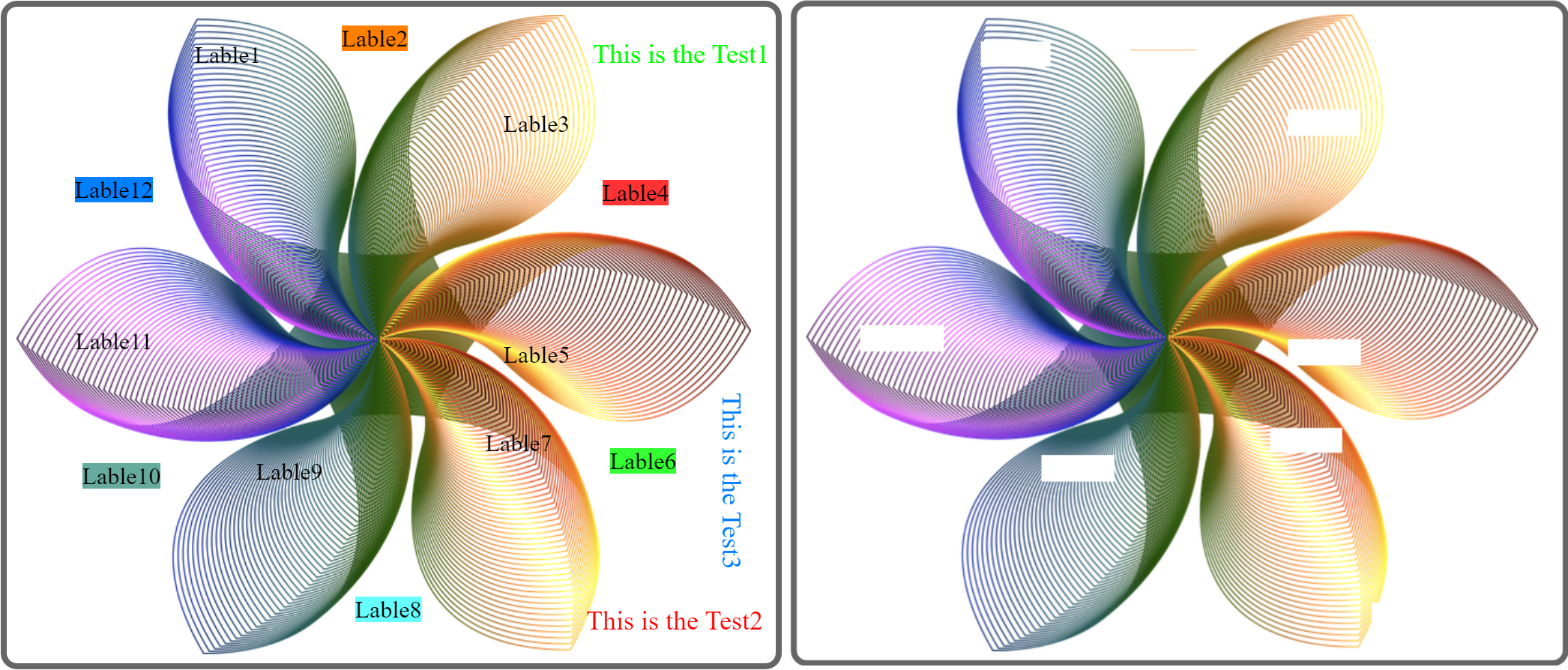}
\caption{Test the TextOCR Practical Performance.}
\label{fig12}
\end{figure}

The GPA $TexOverPic$ and $TextOCR$ are two auxiliary components of LIVE. The main function of GPA $TexOverPic$ is generating the LaTex code that can write some text, especially the citation text over the specific image. The main function of GPA $TextOCR$ is using the Optical Character Recognition (OCR) technologies to recognize the text areas of an image and change the corresponding text areas. As shown in Figure \ref{fig11}, the main principle of $TexOverPic$ is using the LaTex command items $overpic$ and $put$ to generate the LaTex code. The main principle of GPA $TextOCR$ is using $easyocr.Reader(self.languages)$ to recognize the specific areas of the text on the analyzed image and changing the areas to be a pure color area that usually is a white block area. As shown in Figure \ref{fig12}, the recognize-change-performance of $TextOCR$  is tested high.

\section{Evaluation and Conclusion}

LIVE is tested to be useful and meaningful for the interactive LaTeX PDF component design and editing. We proposed a novel design idea to make a LIVE item. More detailed technology and design can be gained from the code project of LIVE. Based on the citation component of PDF and LaTex editing, more LIVE Gitem, GAA or GPA can be developed and designed. Using the LIVE items can express more interactive meaning. Meanwhile, LIVE is easy to develop and prolongable in designing more advanced functions.






\bibliographystyle{IEEEtran}
\bibliography{ref}{}
\end{document}